\documentstyle[12pt]{article}
\begin{document}

\font\rmocho=cmr8
\font\amsit=cmsy10

\newcommand{\ah}{\hbox{\bf a}}
\newcommand{\Ah}{\hbox{\bf A}}
\newcommand{\Bh}{\hbox{\bf B}}
\newcommand{\cero}{\hbox{\bf 0}}
\newcommand{\Dh}{\hbox{\bf D}}
\newcommand{\Eh}{\hbox{\bf E}}
\newcommand{\Fh}{\hbox{\bf F}}
\newcommand{\fh}{\hbox{\bf f}}
\newcommand{\gh}{\hbox{\bf g}}
\newcommand{\Gh}{\hbox{\bf G}}
\newcommand{\Hh}{\hbox{\bf H}}
\newcommand{\Lh}{\hbox{\bf L}}
\newcommand{\Ait}{\hbox{\amsit\char65}}
\newcommand{\Lit}{\hbox{\amsit\char76}}
\newcommand{\Sit}{\hbox{\amsit\char83}}
\newcommand{\mh}{\hbox{\bf m}}
\newcommand{\Mh}{\hbox{\bf M}}
\newcommand{\Oh}{\hbox{\bf O}}
\newcommand{\Ph}{\hbox{\bf P}}
\newcommand{\Rh}{\hbox{\bf R}}
\newcommand{\Sh}{\hbox{\bf S}}
\newcommand{\xh}{\hbox{\bf x}}
\newcommand{\yh}{\hbox{\bf y}}
\newcommand{\zh}{\hbox{\bf z}}
\newcommand{\uno}{\hbox{\bf 1}}
\newcommand{\dpslash}{\partial\kern-7pt /}
\newcommand{\Dhsl}{\Dh\kern-8pt /}
\newcommand{\Phsl}{\Ph\kern-8.6pt /}
\newcommand{\psl}{p \kern-4.5pt /}


\title{\vskip1cm
A nonlocal discretization of fields
\vskip.5cm}
\author{Rafael G. Campos and Eduardo S. Tututi\\ 
Escuela de Ciencias F\'{\i}sico--Matem\'aticas, \\
Universidad Michoacana \\
58060 Morelia, Michoac\'an, M\'exico\\
E-mail: rcampos@zeus.umich.mx; tututi@zeus.umich.mx
\bigskip
\and 
L. O. Pimentel\\ 
Departamento de F\'{\i}sica, \\
Universidad Aut\'onoma Metropolitana--Iztapalapa \\
Apdo. Postal 55-534, M\'exico DF, 09340 M\'exico \\
E-mail: lopr@xanum.uam.mx
}
\date{}
\maketitle
{
\noindent PACS: 11.15.Ha, 11.15.Tk, 02.60.Jh\\
\noindent Keywords: Nonlocal lattices, discrete differentiation, 
discrete fields.}\\

\vspace*{1.5truecm}
\begin{center} Abstract \end{center}
A nonlocal method to obtain discrete classical fields is 
presented. This technique relies on well-behaved matrix representations 
of the derivatives constructed on a non--equispaced lattice. The drawbacks 
of lattice theory like the fermion doubling or the breaking of chiral 
symmetry for the massless case, are absent in this method. 

\vfill
\newpage

\section{Introduction}
The non-perturbative implementation of QCD uses the Feynman 
path--integral approach to yield a discretized version of the theory 
on a equispaced discrete space-time lattice where the main properties 
of the quantum fields are expected to be preserved \cite{Wil74}.  
However, properties such as chirality and uniqueness
of lattice fields are not preserved. In fact, the Nielsen-Ninomiya 
theorem \cite{Nie81} states that for a local and translationally 
invariant hermitian discretization of the fields it is not possible 
to have simultaneously chiral symmetry and uniqueness (no doublers).
Many attempts have been addressed to circumvent the restrictions of 
this no-go theorem (see for example \cite{Sha96}). Among these, one 
can find the use of higher number of dimensions \cite{Kap92}, the 
construction of Dirac operators 
satisfying the Ginsparg--Wilson relation \cite{Gin82}, the use of 
different lattice spacings for the gauge and the fermion fields \cite{Bod96},
or the proposal of nonlocal actions (see for example \cite{Bic00} and references
there in). The locality of the Dirac operator is in general a
desired property, however, there are examples of operators that are 
topologically proper but not local \cite{Tin00}. In this letter 
we introduce another approach to the problem of discretization of fields.
This is done by using a non--equispaced lattice, a projection of the partial 
derivatives that have been used to discretize the classical free Klein-Gordon 
and Dirac fields \cite{Cam00a} and a discrete Fourier transform \cite{Cam92}. 
The classical discrete propagators
approach to their continuum forms when the number of nodes tends to infinity 
and a relationship between this technique and that of path--integrals can be
established. Such a technique can be seen as a projection of the quantum algebras
on a finite linear space yielding matrix representations for the partial
derivatives that produce hermitian actions with nonlocal kinetic terms
\cite{Cam00a, Cam00b}. 
This scheme yields Dirac operators that are free from some 
drawbacks exhibited in the standard lattice method as the fermion doubling, 
the chiral symmetry breaking for the massless case or the yielding of different 
dispersion relations for bosons and fermions.
\\


\section{Discrete formalism}
In order to be concise, we point out in this section the main features of 
the technique and leave the proofs of our assertions in the 
references \cite{Cam00a, Cam92, Cam00b, Cam00c, Cam97}.
\\
We begin by choosing a non--equispaced lattice constructed with 
the set of $N_{\mu}$ ($\mu=0,1,2,3$) zeros $x^{\mu}_j$ of the 
Hermite polynomial $H_{N_{\mu}}(\xi)$ along the direction $\mu$. 
This gives a total of $N=N_0N_1N_2N_3$ lattice points in a non--equispaced 
hypercubic structure. Now, we construct the four $N\times N$ matrices 
\begin{eqnarray}
{\Dh}_0=D_0\otimes 1_{N_3}\otimes 1_{N_2}\otimes 1_{N_1}, \nonumber \\
{\Dh}_3=1_{N_0}\otimes D_3\otimes 1_{N_2}\otimes 1_{N_1}, \label{uno} \\
{\Dh}_2=1_{N_0}\otimes 1_{N_3}\otimes D_2\otimes 1_{N_1}, \nonumber \\
{\Dh}_1=1_{N_0}\otimes 1_{N_3}\otimes 1_{N_2}\otimes D_1, \nonumber
\end{eqnarray}
where $1_{N_\mu}$ is the identity matrix of dimension $N_\mu$ and
$D_\mu$ is the skew-symmetric matrix 
\[
(D_\mu)_{jk}=\cases{0,&{$i=j$},\cr\noalign{\vskip .5truecm}
\displaystyle {1\over{x^\mu_j-x^\mu_k}}, &{$i\not=j$}.\cr} 
\]
We also need 
\[
{\Sh}=S_0\otimes S_3\otimes S_2\otimes S_1 
\]
where $S_\mu$ is a diagonal matrix whose $j$th nonzero element is given 
by the value of the function $\exp(-\xi^2/2)H'_{N_{\mu}}(\xi)$ at 
$\xi=x^{\mu}_j$.
\\
Then, 
\begin{equation}
{\Sh}{\Dh}_\mu{\Sh}^{-1} \label{tres1}
\end{equation} 
is a projection of the partial derivative $\partial_\mu$ in 
the subspace of functions $U$ generated by products of the form 
$u_n(\xi)=\exp(-\xi^2/2)H_n(\xi)$ with $\xi=x^\mu$, $\mu=0,1,2,3$ 
and $n=0,1,\ldots,N_{\mu}-1$. Note that $\{u_n(\xi)\}_0^\infty$, is a 
orthogonal basis for square integrable functions in one dimension. Thus, 
if $\psi(x)\in U$ and $\Psi$ denotes the $N\times 1$ vector of components
\[
\Psi_q=\psi(x_q)\equiv\psi(x^1_j,x^2_k,x^3_l,x^0_m),
\]
ordered according to
\begin{equation}
q=j + (k-1)N_1 + (l-1)N_1N_2 + (m-1)N_1N_2N_3 \label{cuatro}
\end{equation}
where $j=1,2\ldots,N_1$ runs faster than $k=1,2\ldots,N_2$, running
faster than $l=1,2\ldots,N_3$, running faster than $m=1,2\ldots,N_0$,
we have that the vector $\Psi_{,\mu}$ constructed with the values of 
$\psi_{,\mu}=\partial_\mu\psi$ at the site 
$x_q=(x^0_m,x^1_l,x^2_k,x^1_j)$ and ordered as in (\ref{cuatro}), 
is given by
\begin{equation}
\Psi_{,\mu}={\Sh}{\Dh}_\mu{\Sh}^{-1}\Psi. \label{cinco}
\end{equation} 
This means that (\ref{tres1}) 
can be used to reproduce {\it exactly} the values of the partial 
derivatives of functions belonging to $U$ at the lattice sites. 
On the other side, if $\psi(x)$ is a square integrable function 
(considered as a function of just one variable $x^\mu$) that is not 
contained completely in $U$, Eq. (\ref{cinco}) yields approximated
values for the derivatives of $\psi(x)$ at the lattice points. The
error can be estimated in very special cases \cite{Cam97} and become, 
as expected, related to the complement of $\psi(x)$ 
with respect to $U$. This behavior can be shown by considering a 
function of the form $\varphi(x)=a(x)\psi(x)$ with $\psi\in U$ (we 
will take just one variable in the following), 
and discretizing its derivative at $N$ nodes. 
Note that $\varphi$ is not necessarily contained in $U$ and
in general, the vector $\Phi'$ constructed with the values
\begin{equation} 
\varphi'(x_j)=a(x_j)\psi'(x_j)+a'(x_j)\psi(x_j), \label{vphip}
\end{equation}
is not given by $SDS^{-1}\Phi$, where $\Phi$ denotes the
vector constructed with the values $\varphi(x_j)$. Thus, we address 
our attention to the residual vector $R=\Phi'-SDS^{-1}\Phi$. Because
of (\ref{vphip}), $\Phi'$ can be written as $ASDS^{-1}\Psi+A'\Psi$ 
where $A$ is the diagonal matrix containing $a(x_j)$ along the main 
diagonal and $A'$ the diagonal matrix containing $a'(x_j)$. Thus, 
the remaining vector giving the error becomes
\[
R=(ASDS^{-1}-SDS^{-1}A+A')\Psi=S\Delta S^{-1}\Psi, 
\]
where $\Delta$ is the matrix given by $\Delta=[A,D]+A'$, i.e., 
\[ 
\Delta_{jk}=\cases{a'(x_j),&{$j=k$},\cr\noalign{\vskip .5truecm}
\displaystyle{{a(x_j)-a(x_k)}\over{x_j-x_k}}, &{$j\not=k$}.\cr}
\]
By rewriting the commutator as
\begin{equation}
DA=AD+A'-\Delta, \label{da1}
\end{equation}
we can see the way in which the derivative of the product of functions 
$a(x)\psi(x)$ is handled by our method: just construct the diagonal 
matrix with nonzero elements $a(x_j)$ and apply the product $SDAS^{-1}$ 
to the vector $\Psi$. The error $R$ can be determined by considering 
the action of $\Delta$ on an arbitrary vector $V$. To this end, we 
take first the case where $a(x)$ is taken as the power $x^n$. It 
is not difficult to show 
that in such a case, $\Delta V$ is given by the linear combination
\begin{equation}
\Delta V=b_0\xi_{n-1}+b_1\xi_{n-2}+\cdots+b_{n-2}\xi+b_{n-1},
\label{DelV}
\end{equation}
where $\xi_l$ is the vector whose $j$th component is $x^l_j$ and the 
coefficients are given by $b_l=\sum^N_{j=1}x^l_j V_j$. 
The same procedure to get (\ref{DelV}) 
and the expansion of $a(x)$ in powers of $x$ yields 
again (\ref{DelV}) for the partial sum $a_n(x)=\sum_{k=0}^n a^k_n x^k$, 
but the coefficients change to $b_l=\sum^N_{j=1} A_l(x_j) V_j$, 
where $A_l(x)$ is a polynomial of degree $l$. In any case, $R$ is always
in the subspace spanned by (\ref{DelV}) and determined by the higher 
power of $x$ in $a_n(x)$. Now, let us consider the part of $R$ given 
by $\Delta V$, where $V=S^{-1}\Psi$. Since $\psi(x)\in U$, the 
components of $\Psi$ are of the form $\Psi_j=\exp(-x^2_j/2)f(x_j)$ 
(where $f(x)$ is a polynomial of fixed degree), the coefficients 
of (\ref{DelV}) becomes
\begin{equation}
b_l=\sum^N_{j=1} A_l(x_j)\displaystyle{{f(x_j)}\over{H'_N(x_j)}}, 
\qquad l=0,1,\cdots, n-1,
\label{cb2}
\end{equation}
where $n$ is the higher power of $a_n(x)$. Therefore, $S\Delta S^{-1}$ 
projects $\Psi$ on the subspace spanned by the independent vectors 
constructed with the values of $\exp(-x^2/2)x^{N+l-1}$ at the nodes. 
Thus, the error in the discretized derivative of a function 
$a(x)\psi(x)$ not belonging completely to $U$ depends on the components 
$u_l(x)=\exp(-x^2/2)H_l(x)$ defined by the orthogonal complement of $U$. 
Now, to show the condition under which $b_l=0$, let us take the higher 
power of $A_l(x)$ in (\ref{cb2}). This gives a sum of the form 
\begin{equation}
\sum^N_{j=1}\displaystyle{{x^l_jf(x_j)}\over{H'_N(x_j)}}. \label{cb3}
\end{equation} 
By using the matrix representation of $d/dx$ given in \cite{Cal82} applied
to the vector formed with the values of the function $g(x)=(x-x_j)^{l+1}f(x)$ 
at the nodes, one can show that if $f(x)$ is a polynomial of degree at most
$N-l-2$, the sum given in (\ref{cb3}) is zero for a power of $x_j$ less 
or equal to $l$. Thus, the derivative of
the product $a_n(x)\psi(x)$ can be represented exactly by our discrete scheme
whenever $\psi(x)$ is restricted to be in the subspace generated by the
basis $u_l(x)$, $l=0,1,\cdots,N-n-1$.
In the case of four variables, $a(x)$ can be expanded in terms of the basis 
$u_q(x)=u_m(x^0)u_l(x^3)u_k(x^2)u_j(x^1)$, where 
$u_n(\xi)=\exp(-\xi^2/2)H_n(\xi)$ and (\ref{da1}) becomes 
\begin{equation}
{\Dh}_\mu{\ah}={\ah}{\Dh}_\mu+{\ah}_{,\mu}-{\bf\Delta}_\mu,
\label{da2}
\end{equation}
where the projection matrix ${\bf\Delta}_\mu$ can also be given in 
terms of the one--dimensional matrix ${\Delta}_\mu$ through a similar 
relation to (\ref{uno}). Such a matrix makes a projection
on the orthogonal subspace to the one generated by the basis
$u_q(x)$. Therefore, we have a convergent discretization scheme if $N_\mu$ 
is large enough.
\\
Let us take a look on the similarity transformation of (\ref{cinco})
when $N_\mu\to\infty$. First of all, note that (\ref{cinco}) implies
that ${\Dh}_\mu$ can be considered as a representation of 
$\partial_\mu$ for tranformed vectors $\Phi$ of the form 
$\Phi={\Sh}^{-1}\Psi$. Since 
\[
H'_{N_\mu}(x^\mu_j)\approx \displaystyle{(-1)^{N_\mu+j}{{2N_\mu!}\over
{\Gamma({{N_\mu+1}\over 2})}}}\exp[(x^\mu_j)^2/2],\qquad N_\mu\to\infty,
\]
the $j$th nonzero element of $S_\mu$ takes the asymptotic form
$(-1)^{N_\mu+j}2N_\mu!/\Gamma[(N_\mu+1)/2)]$, therefore,
the $jk$ element of $S_\mu D_\mu S_\mu^{-1}$ differs from the $jk$
element of $D_\mu$ only by $(-1)^{j+k}$. This means that in
the limit $N_\mu\to\infty$, the transformed vector $\Phi={\Sh}^{-1}\Psi$
differs from $\Psi$ only by an alternating change of sign along the
$\mu$--direction. In other words, if we take into account the fact that 
the similarity transformation, given by ${\Sh}$, transforms $\Psi$ into 
itself, except for an alternating change of sign along each direction 
in the asymptotic limit, we always may use ${\Dh}_\mu$ instead 
(\ref{tres1}) as representation of $\partial_\mu$.
\\
The commuting matrices (\ref{uno}) can be diagonalized simultaneously
by the unitary and symmetric matrix
\begin{equation}
{\Fh}=F_0^*\otimes F_3\otimes F_2\otimes F_1 \label{seis},
\end{equation}
($*$ means complex conjugate) where 
\begin{equation}
(F_\mu)_{jk}=\sum_{l=0}^{N_\mu-1} (i)^l\varphi_l(x^\mu_j)\varphi_l(p^\mu_k),
\label{siete}
\end{equation}
and
\[
\varphi_l(\xi)=\sqrt{{{(N_\mu-1)!2^{{N_\mu}-1-l}}\over{N_\mu l!}}}
{{H_l(\xi)}\over{H_{{N_\mu}-1}(\xi)}}.
\]
Here, $p^\mu_j$ is also a root of $H_{N_\mu}(\xi)$ and it represents an 
eigenvalue of the discretized linear momentum. It is possible, on physical 
grounds, to construct (\ref{siete}) as a convergent discrete Fourier 
transform \cite{Cam92}. Denoting by $[\cdots]_k$ the $k$th column of a matrix, 
we have that 
\begin{equation}
[F_\mu]_k \to(-1)^{N_\mu+1}[F_\mu^\dagger]_k,
\label{ftofp}
\end{equation}
as $p^\mu_k\to -p^\mu_k$. Besides ${\Fh}^\dagger={\Fh^{-1}}$, 
the basic properties of the four--dimensional discrete Fourier transform 
(\ref{seis}), are
\begin{equation}
i{\Dh}_0 {\Fh}={\Fh}{\Ph}^0, \qquad -i{\Dh}_j {\Fh}={\Fh}{\Ph}^j 
\label{ocho}
\end{equation}
and
\begin{equation}
\lim_{N\to\infty} {\Fh}_{qq'}=C_N e^{-ip_q\cdot x_{q'}}
\label{fasi}
\end{equation}
where $C_N$ is the product of constants of the form\footnote{In fact,
there is a factor $(-1)^{j+k}$ in the asymptotic form of $(F_\mu)_{jk}$ 
that should be taken into account in $C_{N_\mu}$. However, this factor 
can also be treated along the same lines as in the previous paragraph.}
$C_{N_\mu}=2^{{N_\mu}-3/2}(\Gamma[({N_\mu}+1)/2])^2/{N_\mu}!$, and 
$p_q\cdot x_{q'}=p^0_mx^0_{m'}-p^3_lx^3_{l'}-p^2_kx^2_{k'}-p^1_jx^1_{j'}$.
The order between indexes is given by (\ref{cuatro}). In (\ref{ocho})
${\Ph}^\mu$ is given by tensor products of the form (\ref{uno}) in 
terms of the diagonal matrix $P^\mu$ containing the eigenvalues 
$p^\mu_j$ along the main diagonal.
\\


\section{Discrete Lagrangians}
The path--integral quantization of scalar and spinor fields uses 
the classical Lagrangian of the corresponding field. Thus, our aim 
in this section is to give the rules to obtain a discretized version
of such a Lagrangian and their properties. In order to do this, let 
us take for instance the Lagrangian density for QED:
\begin{equation}
L=\bar{\psi}(i{\dpslash}-e\gamma^\mu A_\mu-m)\psi-
{1\over 4}(\partial_\mu A_\nu-\partial_\nu A_\mu)^2, \label{lqed}
\end{equation}
where $\psi=\psi(x)$ and $A_\mu=A_\mu(x)$ are the spinor and
gauge fields respectively.\\
To get a discrete form of this Lagrangian we choose the four--dimensional
lattice $x_q$ of the previous section, i.e, $N_\mu$ zeros of 
$H_{N_\mu}(\xi)$ as lattice points for each direction $\mu=0,1,2,3$ 
giving thus a total of $N=N_0N_1N_2N_3$ nodes. To maintain 
the relationship between the continuous and discrete field variables, 
let us denote by $\Psi$ the  spinor vector of components 
$\Psi_{aq}=\Psi_a(x_q)$, 
where $a=1,2,3,4$ is the spinor index and $q$ is the lattice index 
ordered as in (\ref{cuatro}). A component of this vector can
be described by only one index $r$ according to the order:
\begin{equation}
r=j + (k-1)N_1 + (l-1)N_1N_2 + (m-1)N_1N_2N_3 + (i-1)N_1N_2N_3N_0
\label{ord2}
\end{equation}
where the slower index is $i=1,2,3,4$. Thus, $\Psi$ is a vector of dimension
$4N$. The gauge vector, denoted by ${\Ait}_\mu$, has the components 
${\Ait}_{\mu q}={\Ait}_\mu(x_q)$ ordered as in 
(\ref{cuatro}), yielding a vector of dimension $N$. The values of the
derivative $\partial_\mu A_\nu$ at the nodes are denoted by 
$({\Ait}_{\nu,\mu})_q$. Therefore, a discretization of (\ref{lqed}) is
\begin{equation}
{\Lit}_q=i\bar{\Psi}_{aq}\gamma^\mu_{ab}(\Psi_{,\mu})_{bq}-
\bar{\Psi}_{aq}\big(e\gamma^\mu_{ab}{\Ait}_{\mu q}+m\delta_{ab}
\big)\Psi_{bq}-
{1\over 4}\big[({\Ait}_{\nu,\mu})_q-({\Ait}_{\mu,\nu})_q\big]^2,
\label{lqedq} 
\end{equation}
where $\bar{\Psi}_{aq}=\Psi^\dagger_{bq}\gamma^0_{ba}$. According to 
our procedure, the derivatives $(\Psi_{,\mu})_{bq}$ and
$({\Ait}_{\nu,\mu})_q$ can be approximated by $({\Dh}_\mu)_{qr}\Psi_{br}$
and $({\Dh}_\mu)_{qr}{\Ait}_{\nu r}$, respectively [${\Dh}_\mu$ is used 
instead of (\ref{tres1})]. Here, $a,b=1,2,3,4$
and $q,r=1,\cdots,N$. Repeated indexes are summed excepting $q$. When 
(\ref{lqedq}) is written in matrix form, the reference to the index $q$ 
disappears and the equation becomes the inner product 
\begin{equation}
{\Sit}=\bar{\Psi}\big[i\gamma^\mu\otimes({\Dh}_\mu+ie\tilde{{\Ait}_\mu})-
m {\bf 1}_{4N}\big]\Psi-{1\over 4}
\big({\Dh}^\mu{\Ait}^\nu-{\Dh}^\nu{\Ait}^\mu\big)^{T}
\big({\Dh}_\mu{\Ait}_\nu-{\Dh}_\nu{\Ait}_\mu\big), \label{sqedd}
\end{equation}
where $\tilde{{\Ait}_\mu}$ is a diagonal matrix with the values 
${\Ait}_{\mu q}$ along the main diagonal, $\bar{\Psi}$ is the 
transposed conjugate vector
\[ 
\bar{\Psi}=[(\gamma^0\otimes {\bf 1}_N)\Psi]^\dagger=
\Psi^\dagger(\gamma^0\otimes {\bf 1}_N), 
\]
${\bf 1}_n$ is the identity matrix of dimension $n$ and the superscript $T$ 
means transpose. Because (\ref{sqedd}) involves the sum over all $q$ and 
this means to sum over the spatial and temporal indexes, ${\Sit}$ is 
a discretization of the classical action save for a constant which gives
the correct dimension of the action.\footnote{Indeed, we are using 
dimensionless quantities for both the sites at the lattice (space--time 
and momenta) and the discrete fields.} Since ${\Dh}_\mu$ is real and 
skew--symmetric, ${\Sit}$ is real. By making the definitions
\begin{equation}
\tilde{\Dh}_\mu={\Dh}_\mu+ie\tilde{{\Ait}_\mu} \label{dc}
\end{equation}
and
\[
{\Sit}_M=-{1\over 4}
\big({\Dh}^\mu{\Ait}^\nu-{\Dh}^\nu{\Ait}^\mu\big)^{T}
\big({\Dh}_\mu{\Ait}_\nu-{\Dh}_\nu{\Ait}_\mu\big), 
\]
${\Sit}$ can be written in a simple form
\begin{equation}
{\Sit}=\bar{\Psi}(i\tilde{\Dhsl}-m {\bf 1}_{4N})\Psi+{\Sit}_M, \label{sds}
\end{equation}
with the understanding that the slash notation involves tensor products 
in the discrete case.
\\
\subsection{Chiral symmetry and absence of fermion doubling}
As we mentioned it, our discretization scheme presents some desirable
features to be considered in studies of field theories such as QCD. 
First of all we want to show that the doublers are absent in this 
scheme. Let us begin with the free Dirac equation,
\begin{equation}
(i{\dpslash}-m)\psi=0. \label{eqdl}
\end{equation}
According to our method, the discretized version of this equation is
\begin{equation}
(i\gamma^\mu\otimes {\Dh}_\mu-m {\bf 1}_{4N})\Psi=0. \label{eqdld}
\end{equation}
This is an eigenvalue equation for $m$ and it can be solved by using the
discrete Fourier transform ${\Fh}$ given by (\ref{seis}). Denoting by 
${\Mh}$ the diagonal matrix containing the eigenvalues $m_q$, (\ref{eqdld}) 
becomes
\[
(i\gamma^\mu\otimes {\Dh}_\mu){\bf\Psi}={\bf\Psi}{\Mh}, 
\]
where ${\bf\Psi}$ is the matrix whose $q$th column corresponds to 
the eigenvector $\Psi_q$. Multiplying this equation by 
${\bf 1}_4\otimes{\Fh}^\dagger$ 
and using Eqs. (\ref{ocho}) we obtain the transformed equation
\begin{equation}
(i\gamma^\mu\otimes {\Fh}^\dagger{\Dh}_\mu {\Fh}){\bf\Phi}=
(\gamma^\mu\otimes {\Ph}_\mu){\bf\Phi}={\bf\Phi}{\Mh}, \label{eqdldmt}
\end{equation}
where 
\begin{equation}
{\bf\Phi}=({\bf 1}_4\otimes{\Fh}^\dagger){\bf\Psi} \label{phit}
\end{equation}
is the matrix of transformed eigenvectors and 
${\Ph}_\mu=g_{\mu\nu}{\Ph}^\nu$. This equation, the discrete 
version of ${\psl}\psi=m\psi$, says that ${\Mh}$ is similar to the 
matrix ${\Phsl}$ defined by
\[
{\Phsl}=\gamma^\mu\otimes {\Ph}_\mu.
\]
By applying ${\Phsl}$ again to (\ref{eqdldmt}), one can shows that
${\Mh}^2$ becomes diagonal and equal to ${\bf 1}_4\otimes {\mh}^2$, where 
\[
{\mh}^2=(\Ph^0)^2-(\Ph^1)^2-(\Ph^2)^2-(\Ph^3)^2. 
\]
Thus, the square of a mass eigenvalue, say $m_r$, is given by
the discrete form of the energy--momentum relation 
\begin{equation}
m_r^2=(p^0_m)^2-(p^1_j)^2-(p^2_k)^2-(p^3_l)^2, \label{mr2}
\end{equation}
where the indexes are ordered according to (\ref{ord2}). Therefore,
the doublers are absent. 
\\
Let us note that (\ref{mr2}) yields a dense set of points $m_r^2$ in 
the set of real numbers since $p^\mu_n$ are zeros of the Hermite 
polynomial $H_{N_\mu}(\xi)$. 
\\
In the case where there is a gauge field coupled to a spinor field, 
the discretized Dirac equation is
\[
i\tilde{\Dhsl}\,\Psi=m {\bf 1}_{4N}\Psi. 
\]
This equation defines the Dirac operator as $i\tilde{\Dhsl}$. Due to
the properties of tensor products, this matrix 
satisfies trivially the discretized condition for chiral symmetry:
\[
(\gamma^5\otimes {\bf 1}_N)\,(i\tilde{\Dhsl})+
(i\tilde{\Dhsl})\,(\gamma^5\otimes {\bf 1}_N)=0.
\]
\\
\subsection{Plane--wave solutions for the free Dirac field}
We can obtain solutions of the free equation (\ref{eqdld}) by
anti--transforming (\ref{phit}). Let $\Phi_q$ be the eigenvector 
corresponding to $m_q$ (a vector of zeros everywhere save for 
the $q$th element which is one). Since $m_q$ is degenerate, i.e.,
there are several values of $p^\mu_q$ giving the same eigenvalue $m_r^2$,
the solution of (\ref{eqdld}) $\Psi_{m_r}$ corresponding to the mass 
eigenvalue $m_r$ is given by the linear combination
\begin{equation}
\Psi_{m_r}=\sum_{p^2_q=m^2_r}{\bf c}_q\otimes[{\Fh}]_q,
\label{sol1}
\end{equation}
where $p^2_q=(p^\mu)_q (p_\mu)_q$ and ${\bf c}_q$ is $1\times 4$ vector. 
According to the asymptotic form of the elements of 
$[{\Fh}]_q$ given in (\ref{fasi}), (\ref{sol1}) corresponds to a 
plane--wave discretized solution of (\ref{eqdl}). 
Since (\ref{mr2}) gives the same value $m^2_r$ under the interchange 
$p^\mu_q\to -p^\mu_q$ and $[{\Fh}]_q$ changes to $[{\Fh}^\dagger]_q$, 
except for a minus sign [cf. (\ref{ftofp})] that 
can be included in the linear combination, Eq. 
(\ref{sol1}) splits in two sums, one corresponding to positive values 
of $p^0_m$ and the other to negative values, i.e., 
\begin{equation}
\Psi_{m_r}=\mathop{{\sum}^+}\limits_{p^2_q=m^2_r}{\bf a}_q\otimes[{\Fh}]_q+
\mathop{{\sum}^-}\limits_{p^2_q=m^2_r}{\bf b}_q\otimes[{\Fh}^\dagger]_q,
\label{sol2}
\end{equation}
where the $+$ ($-$) superscript means to sum over points lying on the 
upper (lower) hyperboloids defined through 
\begin{equation}
p^0_m=\pm [m_r^2+(p^1_j)^2+(p^2_k)^2+(p^3_l)^2]^{1/2}. \label{pom}
\end{equation}
To find out the explicit forms of ${\bf a}_q$ and ${\bf b}_q$, let us 
take just one summand of each sum of (\ref{sol2}) and define them as
\[
\Psi^+_q={\bf a}_q\otimes[{\Fh}]_q, \qquad 
\Psi^-_q={\bf b}_q\otimes[{\Fh}^\dagger]_q.
\]
By applying (\ref{eqdld}) to $\Psi^\pm_q$ and taking into account that 
${\Dh}_\mu[{\Fh}]_q=-ip_{\mu q}[{\Fh}]_q$ and 
${\Dh}_\mu[{\Fh}^\dagger]_q=ip_{\mu q}[{\Fh}^\dagger]_q$, where 
$p_{\mu q}$ is the $q$th nonzero element of the diagonal matrix 
${\Ph}_\mu$, we obtain that 
\begin{equation}
(\gamma^\mu p_{\mu q}-m_q{\bf 1}_4){\bf a}_q=0, \qquad 
(\gamma^\mu p_{\mu q}+m_q{\bf 1}_4){\bf b}_q=0,
\label{spi1}
\end{equation}
which are the equations for Dirac spinors evaluated at $p_{\mu q}$ 
and $m_q$ (an expected result since the discretization
scheme presented in this letter is not involved with the spinor space). 
Therefore, the standard procedure to find the solutions of 
(\ref{spi1}) can be followed step--by--step yielding the expansion 
of the wave function of a particle of mass $m_r$ in terms of plane waves
\[
\Psi_{m_r}=\mathop{{\sum}^+}\limits_{p^2_q=m^2_r}[(m_r/p^0_m)
\sum_{l=1,2}a^l_q{\bf u}^l_q]\otimes[{\Fh}]_q+
\mathop{{\sum}^-}\limits_{p^2_q=m^2_r}[(m_r/p^0_m)\sum_{l=1,2}
b^l_q{\bf v}^l_q]\otimes[{\Fh}^\dagger]_q,
\]
where $a^l_q$ and $b^l_q$ are complex numbers and 
\begin{eqnarray*}
&{\bf u}^1_q&=C_{mr}[1,0,p^3_l/(p^0_m+m_r),p^+_{jk}/(p^0_m+m_r)]^T,\\
&{\bf u}^2_q&=C_{mr}[0,1,p^-_{jk}/(p^0_m+m_r),-p^3_l/(p^0_m+m_r)]^T,\\
&{\bf v}^1_q&=C_{mr}[p^3_l/(p^0_m+m_r),p^+_{jk}/(p^0_m+m_r),1,0]^T, \\
&{\bf v}^2_q&=C_{mr}[p^-_{jk}/(p^0_m+m_r),-p^3_l/(p^0_m+m_r),0,1]^T.
\end{eqnarray*}
Here, $C_{mr}=[(p^0_m+m_r)/2m_r]^{1/2}$, $p^\pm_{jk}=p^1_j\pm ip^2_k$ 
and $j,k,l,m$ are the indexes of the
components of $p^2_q$ satisfying (\ref{pom}) for the given eigenvalue $m_r$.
\subsection{Gauge invariance and the covariant derivative}
Let us consider now the way in which the discretized action (\ref{sqedd})
changes under the gauge transformation
\[
\psi(x)\to \displaystyle{e^{-ie\alpha(x)}}\psi(x),\qquad 
A_\mu\to A_\mu+\partial_\mu\alpha(x),
\]
which takes the discretized form
\begin{equation}
\Psi\to ({\bf 1}_4\otimes {\Eh})\Psi,\qquad {\Ait}_\mu\to 
{\Ait}_\mu+\Lambda_{,\mu}, \label{gtd}
\end{equation}
where ${\Eh}$ is the diagonal matrix whose elements are given by
$\exp[-ie\alpha(x_q)]\delta_{qr}$ and $\Lambda_{,\mu}$ is the vector 
whose $q$th component is $\partial_\mu\alpha(x_q)$, ordered according to 
(\ref{cuatro}). We address first our attention to the term 
$\bar{\Psi}(i\tilde{\Dhsl}-m {\bf 1}_{4N})\Psi$ of (\ref{sds})
which we denote by ${\Sit}_{DM}$.
Under this transformation, ${\Sit}_{DM}$ takes the form
\[
{\Sit}_{DM}=\bar{\Psi}({\bf 1}_4\otimes {\Eh}^\dagger)
[i\gamma^\mu\otimes({\Dh}_\mu+ie\tilde{\Ait_\mu}+ie\tilde{\Lambda}_{,\mu})
-m {\bf 1}_{4N}]({\bf 1}_4\otimes {\Eh})\Psi,
\]
where $\tilde{\Lambda}_{,\mu}$ is the diagonal matrix whose principal diagonal
is the vector $\Lambda_{,\mu}$. Therefore,
\[
{\Sit}_{DM}=\bar{\Psi}[i\gamma^\mu\otimes({\Eh}^\dagger{\Dh}_\mu{\Eh}+
ie\tilde{{\Ait}_\mu}+ie\tilde{\Lambda}_{,\mu})-m {\bf 1}_{4N}]\Psi.
\]
In this case, (\ref{da2}) takes the form
\[
{\Dh}_\mu{\Eh}={\Eh}{\Dh}_\mu-ie{\Eh}\tilde{\Lambda}_{,\mu}-{\bf\Delta}_\mu
\]
and this gives
\[
{\Sit}_{DM}=\bar{\Psi}(i\tilde{\Dhsl}-m {\bf 1}_{4N})\Psi-
\bar{\Psi}(i\gamma^\mu\otimes{\Eh}^\dagger {\bf\Delta}_\mu)\Psi,
\]
so that, if $N_\mu\to\infty$, the residual vectors vanish yielding 
invariance of ${\Sit}_{DM}$ under (\ref{gtd}).\\
To show the invariance of ${\Sit}_M$, the term of (\ref{sds}) depending 
only on the gauge field, let us write the vector $\Lambda_{,\mu}$ of 
(\ref{gtd}) as $\Lambda_{,\mu}={\Dh}_\mu\Lambda+R_\mu$, 
where $\Lambda$ is the vector of values $\alpha(x_q)$ and $R_\mu$ is the
residual vector. Since $\alpha(x)$ can be expanded in the basis $u_q(x)$
and ${\Dh}_\mu$ is an exact representation of $\partial_\mu$ for these
functions (save for a similarity transformation), $R_\mu$ must vanish
when $N\to\infty$, so that (\ref{gtd}) takes the asymptotic form
\[
\Psi\to ({\bf 1}_4\otimes {\Eh})\Psi,\qquad {\Ait}_\mu\to 
{\Ait}_\mu+{\Dh}_\mu\Lambda. 
\]
Because of $[{\Dh}_\mu,{\Dh}_\nu]=0$, ${\Sit}_M$ becomes invariant 
in such a limit and this completes the proof of gauge invariance of 
(\ref{sds}) when $N_\mu\to\infty$.
\\
Now, we want to show the way in which the commutator of the discretized 
form of the gauge covariant derivative, Eq. (\ref{dc}), is related to 
the electromagnetic field tensor $F_{\mu\nu}$. By applying the commutator 
$[\tilde{\Dh}_\mu,\tilde{\Dh}_\nu]$ to $\Psi$, we get
\[
[\tilde{\Dh}_\mu,\tilde{\Dh}_\nu]\Psi=ie(\tilde{{\Ait}_\nu}_{,\mu}-
\tilde{{\Ait}_\mu}_{,\nu})\Psi+ie({\bf\Delta}_\nu-{\bf\Delta}_\mu)\Psi,
\]
where we have used again (\ref{da2}). Thus, the expression for $F_{\mu\nu}$ 
in terms of the commutator of the covariant derivative can be recovered 
when $N_\mu\to\infty$.
\\


\section{Final remarks}
Two facts about the discretization procedure presented in this letter 
should be noticed.\\ 
\begin{enumerate}
\item Since it is a nonlocal method (the matrix representations of the 
partial derivatives involve values of the field at each node) based on 
a non--equispaced lattice and the finite number of nodes breaks down 
the translational invariance, the Nielsen--Ninomiya theorem 
does not applies in this instance. In fact, this technique 
yields the right results: the chiral symmetry and the dispersion relation are
obtained correctly when the number of nodes $N$ is finite; the quantum 
commutators, the free propagators, the gauge invariance and the commutator 
of the covariant derivative are 
recovered when $N$ tends to infinity. The translational invariance 
is also recovered in such a limit since the nodes becomes dense in
the space--time.
\item Such a technique involves matrices with a well--behaved structure.  
Thus, it could be a good issue to obtain analytical solutions in some cases.

\end{enumerate}

\end{document}